\documentclass[12pt]{article}

\usepackage{amssymb}
\usepackage{subfig}
\usepackage{graphicx}
\usepackage[colorlinks=true,linkcolor=blue,citecolor=blue]{hyperref}
\usepackage{color}

\usepackage{epsfig,amssymb,amsfonts,amsmath,graphicx,cite}
\usepackage{graphicx}

\begin{document}

\title{Time-evolution of quantum systems via a complex nonlinear Riccati equation. I. Conservative systems with time-independent Hamiltonian}

\author{Hans Cruz${}^1$, Dieter Schuch${}^2$, Octavio Casta\~nos${}^2$, Oscar Rosas-Ortiz${}^3$\\
{\footnotesize ${}^1$Instituto de Ciencias Nucleares, Universidad Nacional Aut\'onoma de M\'exico,}\\ 
{\footnotesize Apartado Postal 70-543, 04510 M\'exico DF,   Mexico}\\
{\footnotesize ${}^2$Institut f\"ur Theoretische Physik, Goethe-Universit\"at Frankfurt am Main,}\\ {\footnotesize Max-von-Laue-Str. 1, D-60438 Frankfurt am Main, Germany}\\
{\footnotesize ${}^3$Physics Department, Cinvestav, AP 14-740, 07000
M\'exico DF, Mexico}}

\date{}
\maketitle

\begin{abstract}{\footnotesize The sensitivity of the evolution of quantum uncertainties to the choice of the initial conditions is shown via a complex nonlinear Riccati equation leading to a reformulation of quantum dynamics. This sensitivity is demonstrated for systems with exact analytic solutions with the form of Gaussian wave packets. In particular, one-dimensional conservative systems with at most quadratic Hamiltonians are studied.}
\end{abstract}

\section{Introduction}
\label{Introduction}

The study of time-dependent (TD) phenomena is important in all areas of physics. In the framework of the time-dependent Schr\"odinger equation (TDSE) there are integrable model Hamiltonians which allow for analytical solutions. Examples related to the harmonic oscillator (HO) system are reviewed in \cite{1}. In classical mechanics integrability, for a time independent Hamiltonian with $n$ degrees of freedom, means that in phase space there are $n$ independent constants of motion (see, e.g., \cite{Arn78,Zak91}). 

Of course, for the conservative one-dimensional case the Hamiltonian system is always integrable. For TD Hamiltonians the situation is completely different because in this case the energy is not a conserved quantity and so even one-dimensional problems can be non-integrable \cite{Som94}.

Wave packets (WPs) that propagate without dispersion were initially studied in quantum mechanics by Schr\"odinger~\cite{2}; these are packets of constant width that are usually discussed in contemporary textbooks on the matter and represent an early example of the coherent states introduced by Glauber~\cite{Gla07}, Klauder~\cite{klauder}, and Sudarshan~\cite{Suda}. That different initial conditions can lead to WPs with totally different properties of the width will naturally emerge from our investigation. This last point  is significant as the HO is still essential in the high-precision measurements of systems with weak dispersion~\cite{3}.  

A different form of studying TD problems in quantum mechanics is via their propagators, transforming an (initial) state at time $t_0$ into a (final) state at a later time $t$. Early on in quantum mechanics, Kennard~\cite{4} derived the propagators for the free particle, the motion of a charged particle in a uniform electric field, and the HO. This subject gained renewed interest when Feynman derived the propagators via his path integral formulation of quantum mechanics which has its origins in the classical principle of least action~\cite{5}. Since then an intensive search for propagators of Hamiltonian systems began (see also~\cite{6}). For (also multidimensional) Hamiltonians that are quadratic in position and momentum, together with those Hamiltonians associated with group-theoretical structures, there are established procedures to derive the propagators. As an example, the method of linear TD invariants introduced in \cite{7,8} shall be mentioned because it is related to our approach for describing the time-dependence of quantum uncertainties, so that solving the problem is traced back to solutions of the classical equations of motion. In particular, the study of the parametric oscillator with TD frequency is very useful because of its applicability to quantum optics, plasma physics and gravitational waves~\cite{9,10}. 

In this work, we consider TDSEs with exact analytic Gaussian WP solutions.  This is the first of a series of three papers that integrate our research. It is known that Gaussian WPs exist for any Hamiltonian that is at most bilinear in position and momentum operators. In this case also exact analytic expressions for the time-evolution of quantum uncertainties exist; in particular for the HO, an oscillator with TD frequency $\omega(t)$ and the free motion. In the case of the parametric oscillator with $\omega=\omega(t)$, the Hamiltonian is no longer a constant of motion but the so-called Ermakov invariant still exists \cite{11}~\footnote{There is an English translation of the work of Ermakov in~\cite{19}.}. Moreover, it is closely related to the Wigner function of the system \cite{12}, and also has been thoroughly used to study the parametric oscillator in classical and quantum physics \cite{13,14,15,16}.  In addition, the Ermakov invariant is relevant in studying the time-evolution of WPs associated to an oscillator with TD frequency. For example, this has been used in the determination of the oscillator's squeezing properties when the frequency is a step function of time~\cite{17}. More recently, the Ermakov invariant has been used in a class of waveguide arrays where the refractive indices and second-neighbor couplings define the mass and the frequency of an analog parametric oscillator~\cite{18}.

In section \ref{section-2}, the TDSEs for conservative systems with at most quadratic potential. Particularly the HO with time-independent frequency $\omega_0$ are considered where the free motion can be obtained in the limit $\omega_0\to 0$. The Gaussian WPs associated to these TDSEs are mainly defined by the coefficient of the quadratic term in the exponential, this coefficient is solution of a complex Riccati equation and  includes information on the quantum uncertainties of position, momentum, and position-momentum correlation. Due to the quadratic term in the Riccati equation, one expects at least two different solutions ($+$ and $-$ signs of the the square root). In addition, as the Riccati equation is a nonlinear (NL) differential equation it is sensitive to the initial conditions and can provide qualitatively quite different behaviour of its solutions depending on the choice of the initial conditions. However, these are related to different physical situations of the initial WP (e.g., minimum uncertainty WP or others). Because of these uncertainties also determine physical properties of the system, like tunnelling currents and ground states energies, these (measurable) properties are also affected by the choice of the initial conditions.

In section \ref{section-3} we show the form in which the initial uncertainties enter the solution of the complex Riccati equation. The relations between the Riccati equation, the Ermakov equation, the Ermakov invariant, and the aforementioned linear dynamical invariants are also established. There it is also shown how the potential (via the Hamiltonian), and thus the TD frequency of the parametric oscillator, can enter the treatment and consequently how our formalism can be extended easily to cases where no analytic solutions may exist for the classical equations of motion. A more detailed discussion will be provided in part III of our study concerning systems with explicitly TD Hamiltonians.

The linearization of the Riccati equation leads to a complex Newtonian equation whose real and imaginary parts are not independent of each other and provide the TD parameters that essentially determine the propagator of the system. They also allow the rewriting of the Ermakov invariant in a form that the comparison with the Wigner function of the system is feasible. This is shown in sections \ref{section-4}, \ref{section-5} and \ref{section-6}.

In the final section, possible generalizations of our method are mentioned, especially the inclusion of dissipative effects which will be discussed in more detail in the part II of our study. There, we shall consider dissipative systems with friction that depends linearly on velocity.


\section{Riccati equations for conservative systems}
\label{section-2}

In the following we consider a one-dimensional oscillator with possibly TD frequency $\omega(t)$. The corresponding TDSE is given by
\begin{equation}
i\hbar\frac{\partial}{\partial t} \Psi(x,t)=\left( -\frac{\hbar^2}{2m}\frac{\partial^2}{\partial x^2} + \frac{m}{2}\omega^2(t) x^2 \right)\Psi(x,t)\,, 
\label{TDSE}
\end{equation}
possessing Gaussian WP solutions of the type
 \begin{equation}
 \Psi(x,t)=N(t)\exp\left\{  i \left[y(t) \tilde{x}^2+\frac{1}{\hbar} \langle p \rangle\tilde{x}+K(t)\right]\right\},
 \label{WP}
 \end{equation}
where $y(t)=y_R(t)+iy_I(t)$ is a TD complex function, $\tilde x= x-\langle x \rangle (t)=x-\eta(t)$, $\eta(t)$ denotes the position of the WP maximum following the classical trajectory and $\langle p \rangle (t)=m\, \dot{\eta}(t)$ is the classical momentum. The purely TD normalizing factor $N(t)$ and phase factor $K(t)$ are not relevant to the following discussion. The imaginary part of $y(t)$ is connected to the WP width, or position uncertainty, via $\sigma^2_x (t) \equiv \langle \tilde{x}^2 \rangle (t)= \frac{1}{4 \, y_I(t)}$. Inserting WP (\ref{WP}) into the TDSE (\ref{TDSE}) one obtains, as equation for the WP maximum\footnote{Depending on the particular time-dependence of $\omega(t)$, the solution of this equation might not even show an oscillatory or periodic behaviour.}, 
\begin{equation}
\ddot{\eta}+\omega^2(t)\eta=0
\label{Newton-equation}
\end{equation}
and the dynamics of the WP width is defined by the complex Riccati equation 
\begin{equation}
\dot{C}+C^2+\omega^2(t)=0
\label{Riccati-equation}
\end{equation}
with $C(t)= \frac{2\hbar}{m}y(t)=C_R(t)+iC_I(t)$ where an overdot denotes derivative with respect to time.

The dynamics of the WP is completely determined by the solutions of the linear Newtonian equation (\ref{Newton-equation}) and the Riccati equation (\ref{Riccati-equation}).

To solve Eq.~(\ref{Riccati-equation}) we use the ansatz $C(t)=\tilde{C}+V(t)$ where $\tilde{C}$ denotes a particular solution of this equation that can be constant or TD. Thus, the inhomogeneous Riccati equation~(\ref{Riccati-equation}) turns into the homogeneous Bernoulli differential equation
\begin{equation}
\dot{V}+2\tilde{C} V+V^2=0 \, .
\label{Bernoulli-equation}
\end{equation}
It is well-known that this differential equation can be linearized via the transformation $V(t)={\kappa(t)}^{-1}$ to yield 
\begin{equation}
\dot{\kappa}-2\tilde{C} \kappa=1.
\end{equation}
The solution of this equation can be given in closed form and depends essentially on the particular solution $\tilde{C}$ and the initial value $\kappa(t_0)\equiv\kappa_0$. For given $\tilde{C}$ and initial condition $\kappa_0$, the solution of the Riccati equation~(\ref{Riccati-equation}) is fixed.

To show the influence of the initial condition $\kappa_0$ we consider, without losing generality, a HO with constant frequency $\omega_0$. In this case, one immediately identifies two particular solutions to the Riccati equation (\ref{Riccati-equation}),
$$
\tilde{C}_\pm=\pm i\omega_0.
$$
The solution $\tilde{C}_{+}=i\omega_0$ leads to the Gaussian WP solution with constant width, usually discussed in connection with the coherent states of the HO in quantum optics textbooks~\footnote{It was found already by Schr\"odinger~\cite{2} as a first example of what is now called a coherent state. The same WPs were retrieved by Glauber~\cite{Gla07}, Klauder~\cite{klauder}, and Sudarshan~\cite{Suda} in their studies on the quantum coherence of light.}. The second solution $\tilde{C}_{-}=-i\omega_0$ would replace the negative sign in the exponent of the Gaussian function by a positive one, leading to a function diverging for $x\to\pm\infty$, i.e., not to a normalizable WP. So, only one of the two possible mathematical solutions of the Riccati equation has physical relevance in this case. It  should be noted that this situation can change when considering the complete solution $C(t)=\tilde{C}_{-}+V(t)$. In this case, a positive imaginary part of $V(t)$ could overcompensate $\tilde{C}_{-}$ and thus allow for a second normalizable WP with TD width. Examples of this kind will be given in Part II of our research, on the topic of dissipative systems. 

For constant $\tilde{C}_\pm$, the analytic solutions of the Bernoulli equation~(\ref{Bernoulli-equation}) take (with $t_0=0$) the form
\begin{equation}
V_{\pm}(t)=\frac{1}{\kappa_{\pm}(t)}=\frac{e^{-2\tilde{C}_\pm t}}{\kappa_0+\frac{1}{2 \tilde{C}_\pm}\left[ 1-e^{-2 \tilde{C}_\pm t} \right]}.
\end{equation}
As can be seen, the functions $V_{\pm}(t)$, and thus $C(t)$, depend strongly on the initial condition $\kappa_0$.  

\begin{figure}[h!]
\begin{center}
\setlength{\unitlength}{1pt}
\begin{picture}(250 ,230)
\includegraphics[width=8cm]{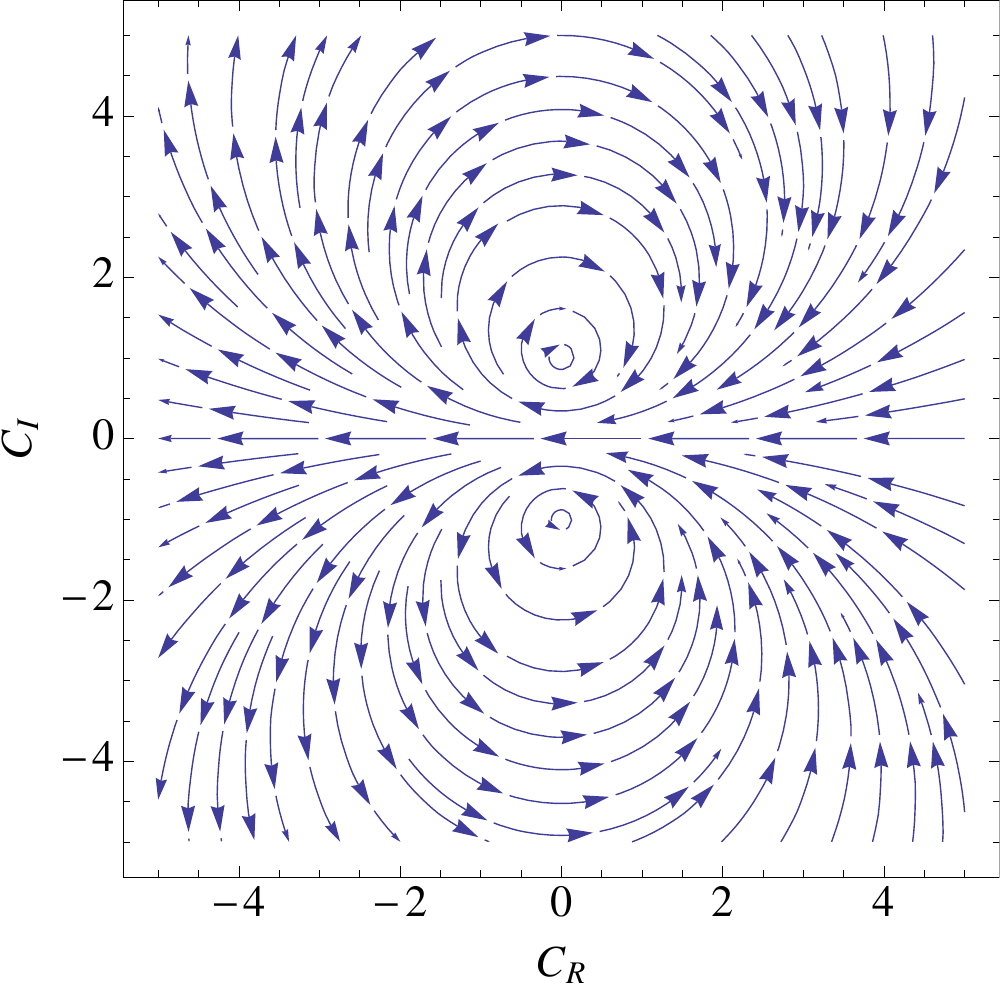}
\end{picture}
\caption{\footnotesize Vectorial field of the Riccati equation for several initial conditions; the frequency is $\omega_0=1$.} \label{riccati-phase-space}
\end{center}
\end{figure}

The vectorial field of the Riccati equation for the HO with constant frequency is displayed in Fig.~\ref{riccati-phase-space}. It has two singular points at $C_\pm=\pm i\omega_0$ (the particular solutions) with periodic orbits around them. Due to the fact that the imaginary part of $C(t)$ is inversely proportional to position uncertainty, all solutions in the lower half-plane are unphysical as they correspond to $C_I(t)< 0$. In addition to the above-mentioned choice of the particular solution, also the value of the initial condition $\kappa_0$ can determine if a solution of the Riccati equation is physical or not.


\section{Riccati--Ermakov connections}
\label{section-3}

The connection of the initial condition $\kappa_0$ to the initial properties of the WP (like its initial width) are demonstrated using a reformulation of the complex Riccati equation in terms of a real 
Ermakov equation. This can be achieved by introducing a new variable $\alpha(t)$ via $C_I(t)=\frac{1}{\alpha^2(t)}$. Inserting this into the complex Riccati equation~(\ref{Riccati-equation}) yields, from its imaginary part, the real part  $C_R(t)=\frac{\dot{\alpha}(t)}{\alpha(t)}$, leading to
\begin{equation}
C(t)=\frac{\dot{\alpha}(t)}{\alpha(t)}+ i \frac{1}{\alpha^2(t)}
\label{Riccati-Ermakov}
\end{equation}
where $\alpha(t)$ is a real function of time and related to the WP width via $\alpha(t)~=~\sqrt{\frac{2m}{\hbar}\sigma^2_x(t)}$. From the real part of the Riccati equation one finally obtains
\begin{equation}
\ddot{\alpha}+\omega^2(t)\alpha=\frac{1}{\alpha^3}.
\label{Ermakov-equation}
\end{equation}
Solving this Ermakov equation is equivalent to solving the Riccati equation. It has been shown by Ermakov~\cite{11,19}
\footnote{The Ermakov equation and the corresponding invariant were reported by Ermakov in 1880 \cite{11}, over the years they have been reinvented several times by various authors \cite{13,20,21}, in a quantum mechanical context, particularly by~\cite{13}. However, the Ermakov system was actually discussed six years earlier in 1874 by Steen~\cite{22}. But, since it was published in Danish in a journal not usually containing articles on mathematics, this work went unnoticed until the end of last century. An English translation of the original paper \cite{23} and generalizations can be found in \cite{24}} 
that by eliminating $\omega$ (TD or time-independent) between equations (\ref{Newton-equation}) and (\ref{Ermakov-equation}) one obtains the dynamical invariant\footnote{Usually, this invariant is given without the factor $\frac{m}{\hbar}$. Factor $m$ came about in view of the fact that Ermakov considered the mathematical equation (\ref{Newton-equation}) whereas the corresponding Newtonian equation in a physical context must be multiplied by $m$. Since the resulting invariant has the dimension of an action, division by $\hbar$ provides a dimensionless constant.}
\begin{equation}
 I=\frac{m}{2\hbar}\left[ (\dot{\eta}\alpha-\eta\dot{\alpha})^2+\left(\frac{\eta}{\alpha}\right)^2 \right].
\label{Ermakov-invariant}
\end{equation}
The quantum uncertainties of position, momentum and their correlation can be determined directly by calculating the related mean values in terms of the Gaussian WP solution. They fulfil a closed system of coupled differential equations (see \ref{Appendix-A}), and can be expressed in terms of $\alpha(t)$ and $\dot{\alpha}(t)$, or, in terms of the real and imaginary parts of $C(t)$, respectively, as:
\begin{eqnarray}
\sigma^2_x(t) & =&\langle \tilde{x}^2 \rangle (t)=\frac{\hbar}{2m} \alpha^2(t)=\frac{\hbar}{2 m} \, \frac{1}{C_I(t)},\label{sigmapos}\\ 
\sigma^2_p(t)&=&\langle \tilde{p}^2 \rangle (t)=\frac{m \hbar}{2} \left[ \dot{\alpha}^2(t)+\frac{1}{\alpha^2(t)} \right]=\frac{m \hbar}{2}  \frac{C^2_R(t) + C^2_I(t)}{C_I(t)} ,\label{sigmamom}  \\
\sigma_{xp}(t)&=&\frac{1}{2}\langle [\tilde{x},\tilde{p}]_{+} \rangle (t)=\frac{1}{2}\langle \tilde{x}\tilde{p}+\tilde{p}\tilde{x} \rangle (t)=\frac{\hbar}{2}\alpha(t) \dot{\alpha}(t)=\frac{\hbar}{2} \frac{C_R(t)}{C_I(t)} \label{sigmacor}
\end{eqnarray}
where Eq.~(\ref{Riccati-Ermakov}) provides the relation between $C(t)$ and $\alpha(t)$. Furthermore, it can be shown straightforwardly that also the Schr\"odinger--Robertson uncertainty relation
$$
\sigma_x^2(t)\sigma_p^2(t)-\sigma^2_{xp}(t)=\frac{\hbar^2}{4}
$$
is fulfilled.

In order to obtain explicit expressions for the time-dependence of the uncertainties,  the Ermakov equation (\ref{Ermakov-equation}) must essentially be solved for given initial conditions $\alpha(t_0)\equiv\alpha_0$ and $\dot{\alpha}(t_0)\equiv\dot{\alpha}_0$ or the Riccati Eq.~ (\ref{Riccati-equation}) for given $\kappa_0$. 

However, it is interesting to note that the solution of the Ermakov equation~(\ref{Ermakov-equation}) can also be constructed knowing two linear independent solutions $\eta_1(t)$ and $\eta_2(t)$ of the Newtonian equation~(\ref{Newton-equation}). This can be achieved using the method of linear invariant operators, introduced by Manko et al.~\cite{7,10}, and their relation with quadratic invariant operators, just as it is outlined in Appendix~\ref{Appendix-B}. The solution of the Ermakov equation~(\ref{Ermakov-equation}) can then be given in the form 
\begin{eqnarray}
\alpha(t)&=&\sqrt{\left( \dot{\alpha}_0^2+\frac{1}{\alpha_0^2}\right)\eta_1^2(t)+\alpha_0^2 \eta_2^2(t)\mp 2\dot{\alpha}_0\alpha_0 \eta_1(t)\eta_2(t) } \,  \nonumber\\
&=&\sqrt{\frac{2m}{\hbar}\left[\sigma^2_{p_0} \eta_1^2(t)+\sigma^2_{x_0} \eta_2^2(t)\mp2\sigma_{xp_0} \eta_1(t) \eta_2(t) \right]}
\label{Solution-Ermakov-equation}
\end{eqnarray}
where the two solutions of the Newton equation and their time-derivatives have the initial conditions
\begin{equation}
\eta_1(t_0)=0, \, \dot{\eta}_1(t_0)=-\frac{1}{m} , \quad
\eta_2(t_0)=1, \, \dot{\eta}_2(t_0)=0. \quad
\label{initial-condition}
\end{equation}
Given Eq.~(\ref{Solution-Ermakov-equation}), the initial conditions of the Ermakov solutions can be expressed in terms of the position and momentum uncertainties together with their correlation function at initial time, i.e.,  
$$
\alpha_{0}= \sqrt{\frac{2m}{\hbar}}\sigma_{x_0},\quad\dot{\alpha}_0=\sqrt{\frac{2}{\hbar m }}\frac{\sigma_{xp_0}}{\sigma_{x_0}} \, .
$$
This is consistent with Eqs. (\ref{sigmapos})-(\ref{sigmacor}) for $t=t_0$:
$$
\sigma^2_x(t_0)=\sigma^2_{x_0},\; \sigma^2_p(t_0)=\sigma^2_{p_0}, \; \sigma_{xp}(t_0)=\sigma_{xp_0} \, .
$$
There is also a relation between the initial condition $\kappa_0$ of the Riccati solution and the initial WP uncertainties. Using (\ref{sigmapos}) and (\ref{sigmacor}) one obtains
\begin{eqnarray*}
C(t_0)\equiv C_0&=&\frac{\dot{\alpha}_0}{\alpha_0}+i \frac{1}{\alpha_0^2}
=\frac{1}{m}\frac{\sigma_{xp_0}}{\sigma^2_{x_0}}+i \frac{\hbar}{2m\sigma^2_{x_0}} \,  
\end{eqnarray*}
so that
$$
V_0=\frac{1}{\kappa_0}=C_0-\tilde{C}_0.
$$
Note that all previous results are valid for TD as well as time-independent frequency $\omega$.

As an example, the HO with constant frequency $\omega_0$ is now considered. For the particular solution $\tilde{C}_{+}=i\omega_0$ we have
$$
V_0=\frac{\dot{\alpha}_0}{\alpha_0}+i\left( \frac{1}{\alpha_0^2}-\omega_0 \right),
\quad
\kappa_0=\frac{\frac{\dot{\alpha}_0}{\alpha_0}-i\left( \frac{1}{\alpha_0^2}-\omega_0 \right)}{\frac{\dot{\alpha}^2_0}{\alpha^2_0}+\left(  \frac{1}{\alpha_0^2}-\omega_0 \right)^2}.
$$
In terms of $\sigma^2_{x_0}$, $\sigma^2_{p_0}$ and $\sigma^2_{xp_0}$, the initial condition $\kappa_0$ can be written as follows
\begin{equation}
\kappa_0=\frac{\frac{1}{2}\sigma_{xp_0}-i\left(\frac{\hbar}{4}-\frac{m\omega_0}{2}\sigma^2_{x_0} \right)}{\frac{1}{2m}\sigma^2_{p_0}+\frac{1}{2}m\omega_0^2\sigma^2_{x_0}-\frac{\hbar\omega_0}{2}}.
\end{equation}

\begin{figure}[ht!]
\begin{center}
\setlength{\unitlength}{1pt}
\begin{picture}(250 ,230)
\includegraphics[width=8cm]{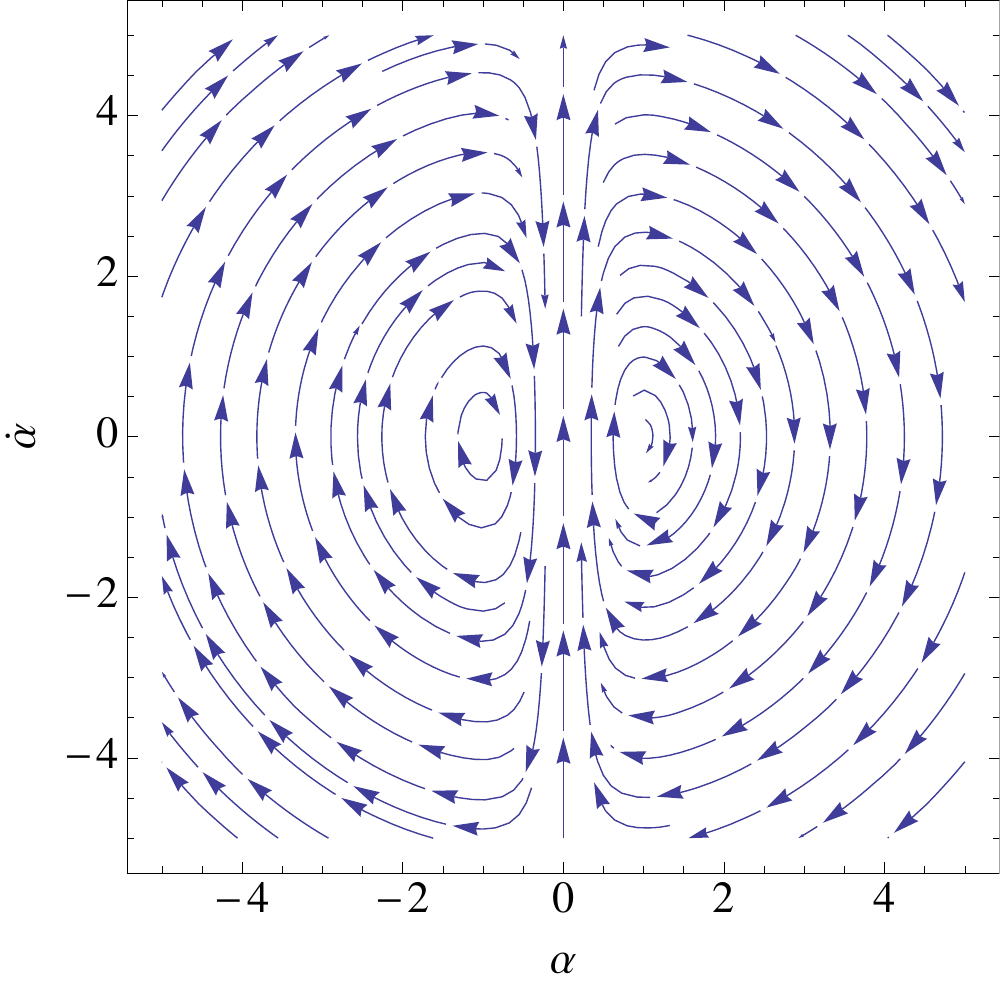}
\end{picture}
\caption{\footnotesize Vectorial field in phase space of the Ermakov equation for the parametric oscillator with constant frequency $\omega_0=1$.}
\label{Ermakov-phase-space}
\end{center}
\end{figure}

The vectorial field in phase space of the Ermakov equation for the HO with constant frequency $\omega_0=1$ is displayed in Fig.~\ref{Ermakov-phase-space}. This has two singular points $\alpha=\pm\frac{1}{\sqrt{\omega_0}}$. The solutions describe orbits around these singular points; the ones on the right correspond to roots of a positive sign whereas the orbits on the left correspond to roots of a negative sign. 

Furthermore, as the integral curves in phase space are closed orbits, the functions $\alpha(t)$ and $\dot{\alpha}(t)$ are periodic. Therefore, the uncertainties and the correlation function are also periodic. An exceptional case is obtained with the singular solutions for which one obtains 
$$
\sigma^2_x(t)=\frac{\hbar}{2m\omega_0},\; \sigma^2_p(t)=\frac{\hbar m \omega_0}{2}, \; \sigma_{xp}(t)=0.
$$ 
These values correspond to the ground state of the HO and to the coherent state with constant width. Therefore, if the initial WP is not this coherent state, the uncertainties and the correlation function evolve periodically in time.

The analytic expression of the solution of the Ermakov equation is
\begin{eqnarray}
\alpha(t)&=\sqrt{\frac{2m}{\hbar}\left[ \frac{\sigma^2_{p_0}}{m^2\omega_0^2} \sin^2{\omega_0 t}+\sigma^2_{x_0}\cos^2{\omega_0 t}\mp\frac{2\sigma_{xp_0}}{m\omega_0}\sin{\omega_0 t}\cos{\omega_0 t} \right]} \, \nonumber \\
&=\alpha_0\sqrt{ \left(\frac{\dot{\alpha}_0^2}{\alpha_0^2}+\frac{1}{\alpha_0^4} \right)\frac{1}{\omega_0^2} \sin^2{\omega_0 t}+\cos^2{\omega_0 t}\mp\frac{2}{\omega_0}\frac{\dot{\alpha}_0}{\alpha_0}\sin{\omega_0 t}\cos{\omega_0 t}},
\label{alpha-HO}
\end{eqnarray}
so that the uncertainties and the correlation functions are
\begin{eqnarray}
\sigma^2_x(t) & =&\frac{\sigma^2_{p_0}}{m^2\omega_0^2} \sin^2{\omega_0 t}+\sigma^2_{x_0}\cos^2{\omega_0 t}+ \frac{2\sigma_{xp_0}}{m\omega_0}\sin{\omega_0 t}\cos{\omega_0 t},
\label{sx-HO} \\ 
\sigma^2_p(t)&=&\sigma^2_{p_0}\cos^2{\omega_0 t}+m^2\omega_0^2\sigma^2_{x_0}\sin^2{\omega_0 t}-2m\omega_0\sigma_{xp_0}\sin{\omega_0 t}\cos{\omega_0 t}, \qquad \\
\sigma_{xp}(t)&=& \left(\frac{\sigma_{p_0}^2}{2m\omega_0}-\frac{m\omega_0\sigma^2_{x_0}}{2} \right)\sin{2\omega_0 t}+\sigma_{xp_0}\cos{2\omega_0 t}.
\label{sxp-HO}
\end{eqnarray}
The corresponding results for the free motion are obtained easily using
\begin{equation} 
\lim_{\omega_0 \to 0} \frac{\sin{\omega_0 t}}{\omega_0} = t, \quad \lim_{\omega_0 \to 0} \cos{\omega_0t} = 1
\end{equation} 
and are presented in Appendix D.

These general expressions lead to particular physical solutions once the initial values $\sigma^2_{x_0}$, $\sigma^2_{p_0}$ and $\sigma_{xp_0}$ (or $\alpha_0$ and $\dot{\alpha}_0$) are given.  

\begin{figure}[ht!]
   \centering
   \subfloat[]{
        \label{fig0:museo:a}         
              \includegraphics[width=0.44\textwidth]{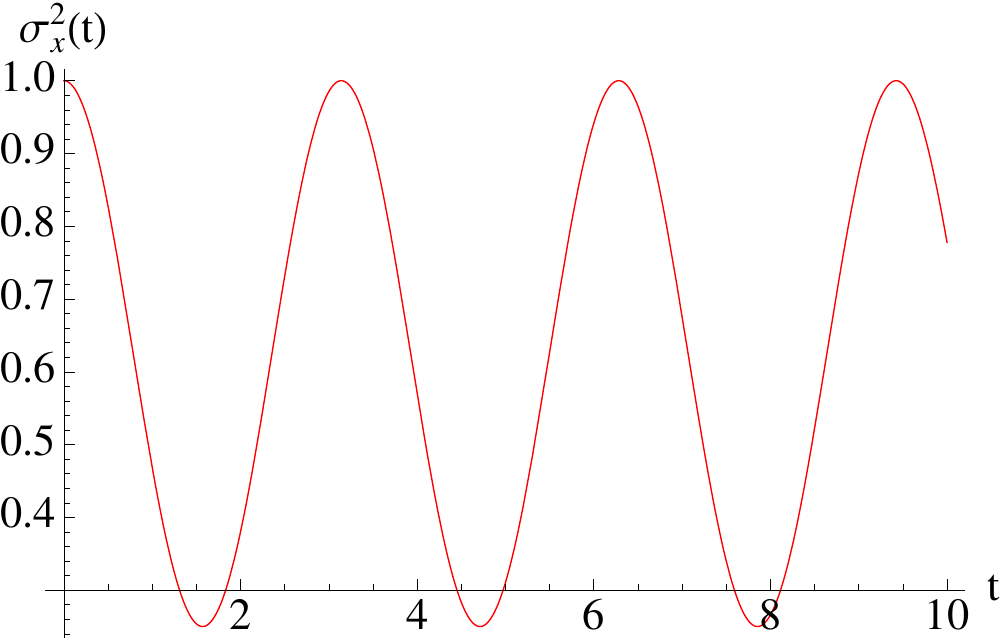}
              }
   \hspace{0.05\linewidth}
   \subfloat[]{
        \label{fig0:museo:b}         
                \includegraphics[width=0.44\textwidth]{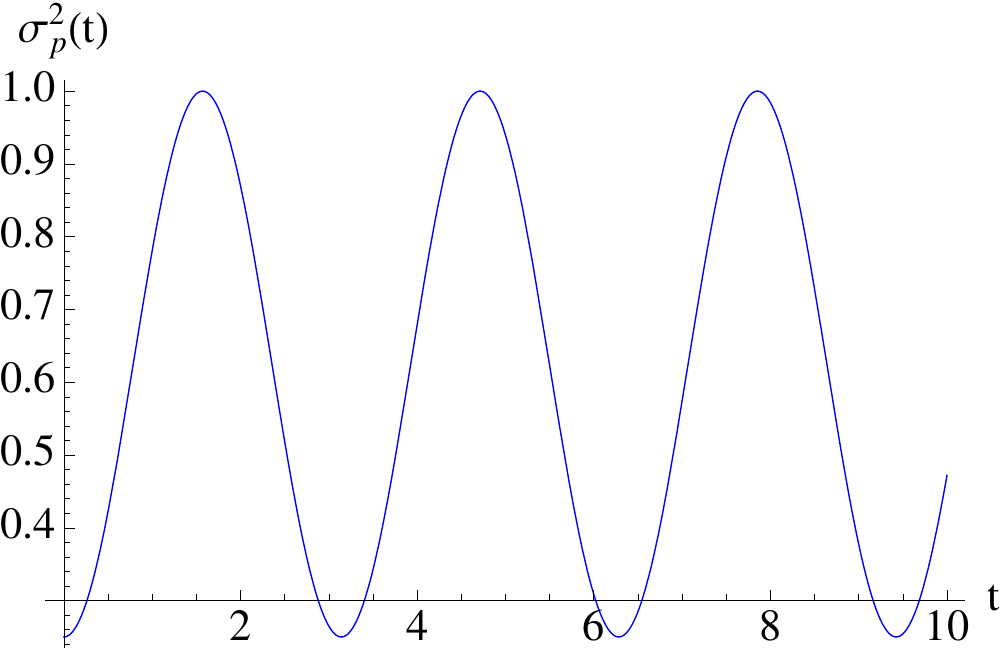}
                }\\
   \subfloat[]{
        \label{fig0:museo:c}         
                \includegraphics[width=0.44\textwidth]{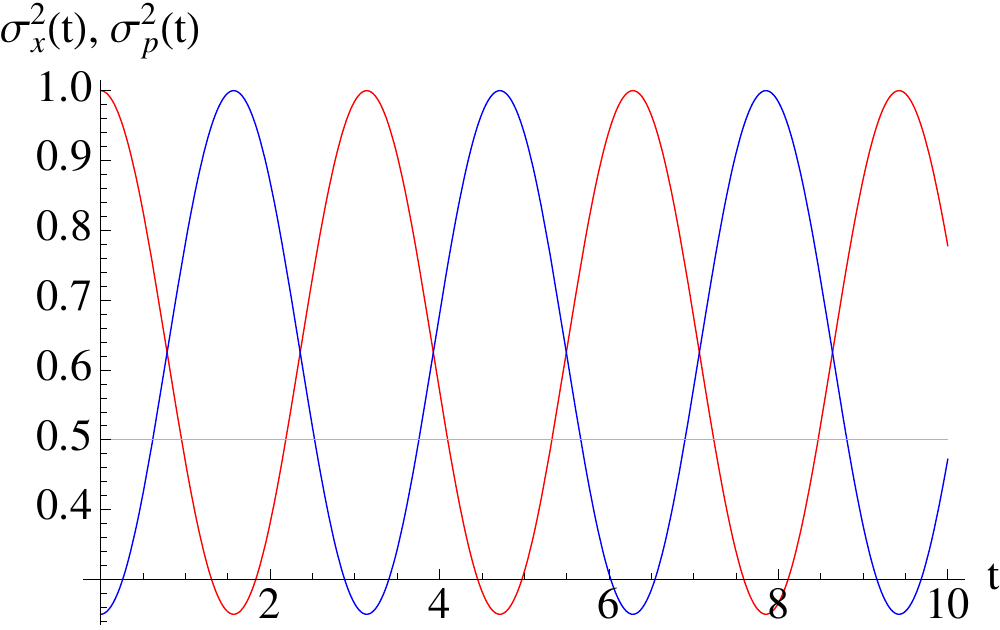}
                }
                 \hspace{0.05\linewidth}
   \subfloat[]{
        \label{fig0:museo:d}         
                \includegraphics[width=0.44\textwidth]{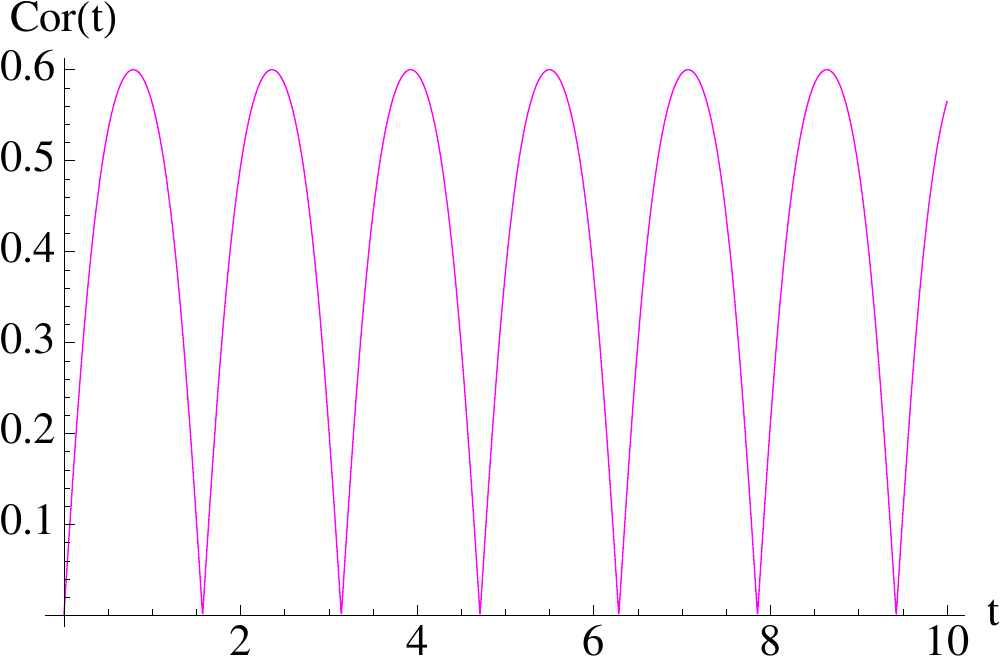}
                }
   \caption{\footnotesize Time-dependence of position (a) and momentum (b) uncertainties. In (c) both uncertainties are plotted together with a straight line in cyan colour determining if the WP is squeezed. In (d) the correlation coefficient $Cor(t)$ is given as a function of time; in all cases we consider $m=\hbar=1$.  }
   	\label{uncertainties}                
\end{figure}

For example, consider a WP with parameter $\alpha_0$ given in terms of constant frequency $\omega_1=0.5$ as $\alpha_0=\frac{1}{\sqrt{\omega_1}}$. This WP corresponds to the coherent state of a HO-Hamiltonian with frequency $\omega_1$ (i.e., $1/\alpha_o^2 \neq \omega_0$) . The initial uncertainties and correlation function are given, respectively, by 
\[
\sigma^2_{x_0}=\frac{\hbar}{2m\omega_1} \, , \quad  \sigma^2_{p_0}=\frac{\hbar m\omega_1}{2} \, ,  \quad \sigma_{xp_0}=0 \, .
\] 

The time-evolution of position and momentum uncertainties, with parameters $m=\hbar=1$, are displayed in Fig.~\ref{fig0:museo:a} and  Fig.~\ref{fig0:museo:b}, respectively. Both uncertainties are periodic functions of time. In Fig.~\ref{fig0:museo:c}, $\sigma_x^2(t)$ and $\sigma_p^2(t)$ are plotted together with a straight line indicating the presence of squeezing of the WP (the uncertainty $\sigma_x^2(t)$ or $\sigma_p^2(t)$ takes a value below this line). In Fig.~\ref{fig0:museo:d} the correlation coefficient, defined as
\begin{equation}
Cor(t)=\frac{|\sigma_{xp}(t)|}{\sigma_x(t)\sigma_p(t)} \, ,
\label{cor-coef}
\end{equation}
is plotted. Note that $Cor(t)$ evolves periodically in time from zero to approximately $0.6$.


\section{Riccati--Newton connections}
\label{section-4}

The Riccati equation (\ref{Riccati-equation}) can be linearized replacing the complex variable $C(t)$ by the logarithmic derivative of another complex variable, $\lambda(t)=\lambda_R(t)+i\lambda_I(t)$, according to
\begin{equation}
C(t)=\frac{\dot{\lambda}(t)}{\lambda(t)}.
\end{equation}
Inserting this into the Riccati equation leads to the complex Newtonian equation
\begin{equation}
\ddot{\lambda}(t)+\omega^2(t)\lambda(t)=0.
\label{Complex-Newton-equation}
\end{equation}
Writing $\lambda(t)$ in polar coordinates as $\lambda(t)=\alpha(t)e^{i\phi(t)}$, the corresponding Riccati variable reads $C(t)=\frac{\dot{\alpha}(t)}{\alpha(t)}+i\dot{\phi}(t)$. This is in agreement with definition (\ref{Riccati-Ermakov}), providing $\dot{\phi}(t)=\frac{1}{\alpha^2(t)}$ holds, which looks similar to the conservation of angular momentum for the motion of a system with central forces. The validity of this conservation law can easily be shown by inserting the above form of $C(t)$ into the Riccati equation and looking at its imaginary part. It also follows from the Wronskian of the complex Newtonian equation,
\begin{equation}
\dot{\lambda}_I(t)\lambda_R(t)-\lambda_I(t)\dot{\lambda}_R(t)=1,
\label{conservation-law}
\end{equation}
if real and imaginary parts of $\lambda(t)$ are written as $\lambda_R(t)=\alpha(t)\cos{\phi(t)}$ and $\lambda_I(t)~=~\alpha(t)\sin{\phi(t)}$. Apparently, one of the components of $\lambda(t)$ is, up to a constant factor $c$, identical to the classical trajectory $\eta(t)$. For reasons that become clear later on, we choose $\lambda_I(t)=c\eta(t)$. Rewriting the Ermakov invariant (\ref{Ermakov-invariant}) in terms of $\lambda_I(t)$ and $\dot{\lambda}_I(t)$ instead of $\eta(t)$ and $\dot{\eta}(t)$, it is obvious that the second term of the invariant reduces to $\left( \frac{\lambda_I(t)}{\alpha(t)}\right)^2=\sin^2{\phi}(t)$. Therefore the first term of the Ermakov invariant must be $\cos^2{\phi(t)}$, i.e.,
\begin{equation}
\frac{\lambda_R^2(t)}{\alpha^2(t)}=\left( \dot{\lambda}_I(t)\alpha(t)-\lambda_I(t)\dot{\alpha}(t) \right)^2=\cos^2{\phi(t)}.
\end{equation}
This expression allows one to determine $\lambda_R(t)$ in closed form up to a $\pm$ sign,
\begin{equation}
\lambda_R(t)=\pm c\alpha^2(t)\left( \dot{\eta}(t)-\frac{\dot{\alpha}(t)}{\alpha(t)}\eta(t) \right)
\label{lambda-real}
\end{equation}
with $c=\sqrt{\frac{m}{2\hbar I}}$.

Using Eqs. (\ref{sigmapos}) and (\ref{sigmacor}), one can write in matrix form
\begin{eqnarray}
\left(
   \begin{array}{c}
      \lambda_R(t)  \\
      \lambda_I (t) \\
   \end{array}
   \right)
   &=& 
    \left(
	 \begin{array}{cc} 
	      \mp c\alpha(t) \dot{\alpha}(t) & \pm \frac{c}{m} \alpha^2(t) \\
	      c & 0 \\
	   \end{array}
	   \right)
              	     \left(
	   	             \begin{array}{c}
			     \langle x \rangle(t) \\
		              \langle p \rangle(t) \\
		               \end{array} 
                      \right)\nonumber\\
   &=&  
	\left(
	 \begin{array}{cc} 
	      \mp\frac{2 c}{\hbar }\sigma_{xp}(t) & \pm \frac{2c}{\hbar } \sigma^2_{x}(t) \\
	      c & 0 \\
	   \end{array}
	   \right)
              	     \left(
	   	             \begin{array}{c}
			     \langle x \rangle(t) \\
		              \langle p \rangle(t) \\
		               \end{array} 
                      \right) \, .
\label{Lambda-classic-solution}
\end{eqnarray}

Hence $\langle x \rangle(t)=\eta(t)$ and $\langle p \rangle(t)=m\dot{\eta}(t)$ are still two linear-independent solutions of the linear second-order differential equation (\ref{Newton-equation}); but they are no longer orthogonal, whereas $\lambda_R(t)$ and $\lambda_I(t)$ are. In Eq. (\ref{Solution-Ermakov-equation}) two linear-independent solutions obeying Eq. (\ref{Newton-equation}) and the initial conditions (\ref{initial-condition}) are needed to obtain the Ermakov variable $\alpha(t)$. One solution can always be the classical trajectory $\langle x \rangle(t)=\eta(t)$ or $\lambda_I(t)$ which differs from $\eta(t)$ only by a constant factor. As a second solution $\lambda_R(t)$ can obviously be chosen. In cases where $\dot{\alpha}_0$ or $\frac{\dot{\alpha}_0}{\alpha_0}=C_{R_0}$ vanish, also $\dot{\eta}(t)$ can be used because, according to (\ref{lambda-real}), $\lambda_R(t)$ and $\dot{\eta}(t)$ then fulfil (up to a constant factor) the same initial conditions. In the cases of the free motion ($V(x,t)=0$) and the HO, with Gaussian WP solutions (also with oscillating width) one actually finds this situation, $\frac{\dot{\alpha}_0}{\alpha_0}=0$.

Rewriting the Ermakov invariant (\ref{Ermakov-invariant}) in terms of $\lambda_R(t)$ and $\lambda_I(t)$ as
\begin{equation}
I=\frac{m}{2\hbar}\frac{1}{c^2}\left[ \left( \frac{\lambda_R}{\alpha} \right)^2+\left( \frac{\lambda_I}{\alpha} \right)^2 \right]
\end{equation}
and using (\ref{Lambda-classic-solution}), it can be expressed in terms of bilinear combinations of the mean value of position $\langle x \rangle(t)=\eta(t)$ and momentum $\langle p \rangle(t)=m\dot{\eta}(t)$ multiplied by the corresponding conjugate uncertainties as
\begin{equation}
I(\langle x 	\rangle, \langle p \rangle, t)=\frac{1}{\hbar^2}\left[ \sigma_p^2(t)\langle x \rangle^2(t)-2\sigma_{xp}(t)\langle x \rangle(t)\langle p \rangle(t)+\sigma_x^2(t)\langle p \rangle^2(t) \right].
\label{Ermakov-invariant-physc}
\end{equation}
Finally, writing $\alpha^2(t)$ as $\alpha^2=\lambda^2_R(t)+\lambda^2_I(t)$ and using (\ref{Lambda-classic-solution}), one obtains an expression that is in agreement with Eq. (\ref{Solution-Ermakov-equation}). 


\section{Feynman kernel/propagator}
\label{section-5}

A different way of describing the time-evolution of a quantum system is to apply a propagator $G(x,x';t,t')$ to an initial state $\Psi(x',t')$ that transforms this state into $\Psi(x,t)$ at position $x$ and time $t$ according to
\begin{equation}
\Psi(x,t)=\int \, dx' G(x,x';t,t')\Psi(x',t'=0). 
\label{kernel}
\end{equation}
The integral kernel $G(x,x';t,t')$ is also called Feynman kernel and was determined by Feynman using his path integral method~\cite{5}.

In the cases discussed in this work, the propagator can be expressed easily in terms of real and imaginary parts of the complex variable $\lambda (t)$ that fulfils the complex Newtonian equation (\ref{Complex-Newton-equation}). Particularly for our case with Gaussian WP solutions of the TDSE, it can be shown that for an initial Gaussian state that can be written as \footnote{If the initial change of the WP width, and thus $\dot{\alpha}_0$, is different from zero, the term $i \left(\frac{x'}{\alpha_0}\right)^2$ in Eq. (\ref{initial WP}) has to be replaced by $\left(\frac{\dot{\alpha}_0}{\alpha_0} + i \frac{1}{\alpha_0^2}\right) x'^2 = C_0 x'^2$. However, in the examples discussed in this paper, this is not the case.}  
\begin{equation}
\Psi(x',t'=0) = \left( \frac{m}{\pi \hbar \alpha_0^2} \right)^{1/4} \exp\left\{ \frac{im}{2\hbar}\left[ i \left(\frac{x'}{\alpha_0}\right)^2 + 2\frac{p_0}{m} x' \right] \right\}, 
\label{initial WP}
\end{equation}
with $p_0 = \langle p \rangle (t_0=0)$, also the propagator has the form of a Gaussian function. The exponent of this function then obviously contains terms proportional to $x^2$, $x'^2$ and $xx'$ with appropriate TD coefficients. These coefficients can be easily expressed in terms of $\lambda_R(t)$ and $\lambda_I(t)$ and the propagator can be written in the form
\begin{eqnarray}
G(x,x';t,t'=0)&=&\left( \frac{m}{2\pi i \hbar \alpha_0 \lambda_I(t)} \right)^{1/2}\exp\Bigg\{ \frac{im}{2\hbar}\Bigg[ \frac{\dot{\lambda}_I(t)}{\lambda_I(t)}x^2-2\frac{x}{\lambda_I(t)}\left(\frac{x'}{\alpha_0}\right)\nonumber\\
&&+ \frac{\lambda_R(t)}{\lambda_I(t)} \left( \frac{x'}{\alpha_0} \right)^2 \Bigg] \Bigg\}.
\label{propagator}
\end{eqnarray}
As the initial WP $\Psi(x',t'=0)$ does not depend on $x$ and $t$, the propagator $G(x,x';t,t')$ itself must also fulfil the TDSE. Inserting $G(x,x';t,t')$ into this equation leads to terms proportional to $x^2$, $x$ and independent of $x$. The coefficient of the latter ones provide the conservation law (\ref{conservation-law}); the coefficient of the terms proportional to $x$ cancel each other. From the ones proportional to $x^2$, one obtains a Riccati equation for $\left( \frac{\dot{\lambda}_I(t)}{\lambda_I(t)} \right)$ as variable (instead of $C(t)= \frac{\dot{\lambda}(t)}{\lambda(t)} $) that can be linearized to the Newtonian equation for $\lambda_I(t)$ (which is proportional to $\eta(t)=\langle x \rangle(t)$).

Inserting the propagator (\ref{propagator}) into Eq. (\ref{kernel}) and performing the integration now leads to a Gaussian WP in terms of $x$ and $\lambda_I(t)$, $\lambda_R(t)$ instead of $x$ and $\langle x \rangle(t)$, $\langle p \rangle(t)$, i.e.,
\begin{eqnarray}
\Psi_{WP}(x,t)&=&\left( \frac{m}{\pi\hbar} \right)^{1/4}\left( \frac{1}{\lambda_R(t)+i\lambda_I(t)} \right)^{1/2}\exp\Bigg\{ \frac{im}{2\hbar}\Bigg[ \frac{\dot{\lambda}_I(t)}{\lambda_I(t)}x^2\nonumber\\
&&-\frac{(x-\frac{1}{c}\lambda_I(t))^2}{\lambda_I(t)(\lambda_R(t)+i\lambda_I(t))} \Bigg] \Bigg\}.
\end{eqnarray}
Comparing this with the WP in the form (\ref{WP}) shows that $\frac{1}{c}\lambda_I(t)=\eta(t)$ justifies our choice made above. With the help of conservation law (\ref{conservation-law}), it is also easy to prove that the coefficient of the terms quadratic in $x$ simply equals $iy(t)=\frac{im}{2\hbar}\left( \frac{2\hbar}{m}y(t) \right)=\frac{im}{2\hbar}C(t)$.


\section{Wigner function}
\label{section-6}

The Wigner function of the systems under consideration can be obtained from the WP solution (\ref{WP}) via the transformation
\begin{equation}
W(x,p;t)=\frac{1}{2\pi\hbar}\int_{-\infty}^{\infty} dq\;\Psi^\ast\left( x+\frac{q}{2},t \right)\Psi\left( x-\frac{q}{2},t \right)e^{\frac{i}{\hbar}qp} \, 
\end{equation}
and can be written as
\begin{equation}
W(x,p;t)=\frac{1}{\pi\hbar}\exp\left\{ -2\left( \frac{y_I^2+y_R^2}{y_I} \right)\tilde{x}^2-\frac{\tilde{p}^2}{2\hbar^2y_I} +\frac{2y_R}{\hbar y_I}\tilde{x}\tilde{p} \right\} \, .
\end{equation}
Using (\ref{Riccati-Ermakov}), (\ref{sigmapos}), (\ref{sigmamom}) and (\ref{sigmacor}), this finally acquires the form
\begin{equation}
W(x,p;t)=\frac{1}{\pi\hbar}\exp\left\{ -\frac{2}{\hbar^2}\left[ \sigma_p^2(t)\tilde{x}^2(t) -2\sigma_{xp}(t)\tilde{x}(t) \tilde{p}(t) +\sigma_x^2(t)\tilde{ p }^2(t) \right] \right\} \, .
\label{Wigner-physic}
\end{equation} 

Comparing this with the Ermakov invariant (\ref{Ermakov-invariant-physc}), one realizes that the Wigner function can be expressed as
$$
W(x,p;t)=\frac{1}{\pi\hbar}e^{ - 2 I(\tilde{x},\tilde{p};t)} \, .
$$
If we consider $\langle x \rangle(t)=0$ and $\langle p \rangle(t)=0$, this is reduced to 
\begin{equation}
W(x,p;t)=\frac{1}{\pi\hbar}e^{ - 2 I(x,p;t)}.
\end{equation}
Thus, when the expectation values of position and momentum are zero for any time, the Ermakov invariant $I(x,p;t)$ explicitly defines the analytic form of the Wigner distribution $W(x,p; t)$. Remarkably, in phase space the invariant $I(x,p;t)$ describes ellipses that are rotating and thus their major and minor axes are changing in time. This behaviour is inherited by the maximum of the Wigner distribution which will describe also elliptical trajectories with the rotating axis directed along its maximum value.

In the most general situation, i.e. $\langle x \rangle(t) \neq 0$ and $\langle p \rangle(t) \neq 0$, the Wigner function displays the same behaviour; however, the maximum of the distribution will move in phase space according to the classical trajectory $(\langle x \rangle(t) ,\langle p \rangle(t) )$.

\begin{figure}[ht!]
\begin{center}
\setlength{\unitlength}{1pt}
\begin{picture}(250 ,210)
\includegraphics[width=8cm]{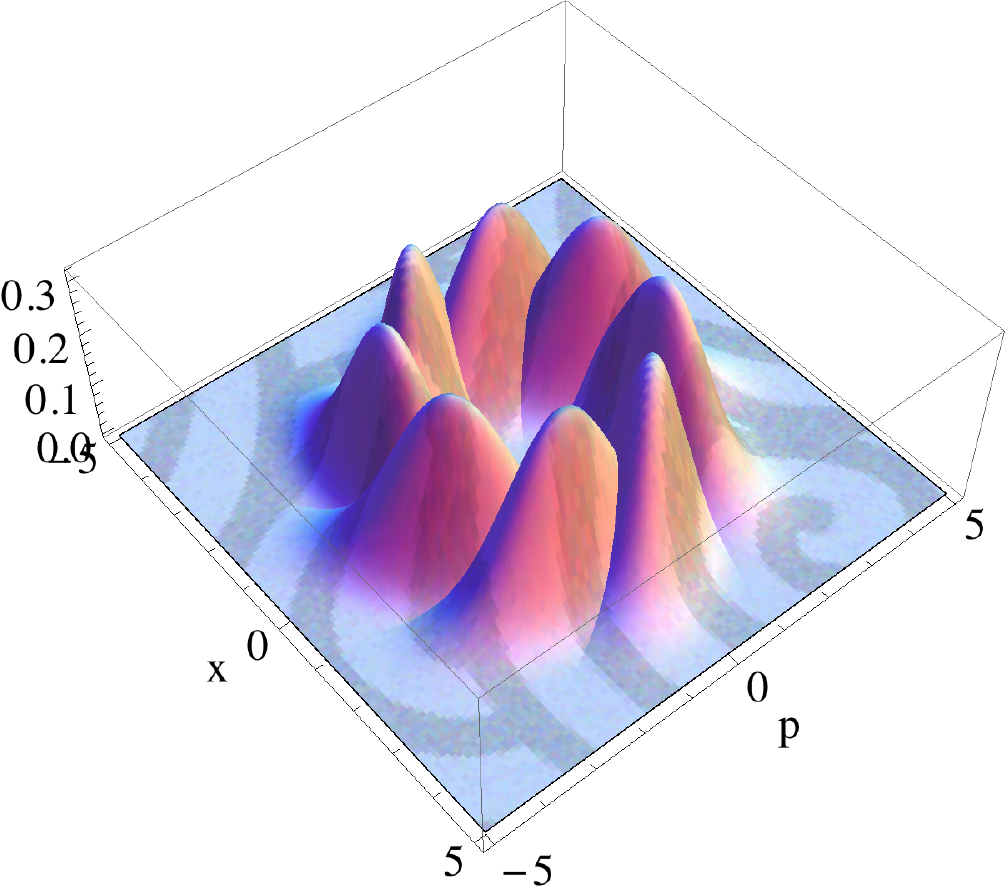}
\end{picture}
\caption{\footnotesize The Wigner function for the parametric oscillator with parameters $m=~\hbar=~\omega_0=~1$ where the initial state is a coherent state with parameter depending on the frequency $\omega_1=0.5$. This implies that $\sigma^2_{x_0}=1$ and $\sigma^2_{p_0}=2$. We also use $\langle x \rangle (t_0=0)=1$ and $\langle p \rangle (t_0=0)=2$.  The times used are $t=\{0, \pi/4, \pi/2, 3 \pi /4,5 \pi/4, 3 \pi/2, 7 \pi/4, 2 \pi \}$.  }
\label{wigner-function}
\end{center}
\end{figure}

As an example, we consider the Wigner function of an initial WP with its parameter depending on the frequency $\omega_1$ and that is associated with a HO with constant frequency $\omega_1$. Thus, the initial uncertainties and correlation function are given by $\sigma^2_{x_0}=\frac{\hbar}{2m\omega_1}$, $\sigma^2_{p_0}=\frac{\hbar m \omega_1}{2}$ and $\sigma_{xp_0}=0$, respectively. The related Wigner function is plotted in Fig.~\ref{wigner-function} for different times where we have used the parameters $m=\hbar=\omega_0=1$, $\omega_1=0.5$ and the initial expectation values $\langle x \rangle (t_0=0)=1$ and $\langle p \rangle (t_0=0)=2$. Notice that the maximum of the Wigner function describes an elliptic trajectory that is defined by the classical energy of the system
\begin{equation}
\frac{\langle p \rangle^2(t)}{2m}+\frac{1}{2}m\omega_0^2\langle x \rangle^2(t)=E_{cl}.
\end{equation}
Furthermore, the probability distribution functions in position and momentum representation can be obtained via
\begin{eqnarray}
\Psi^\ast(x,t)\Psi(x,t)&=&|\Psi(x,t)|^2=\int^{\infty}_{-\infty} dp\; W(x,p;t)  \, ,\\ 
\Psi^\ast(p,t)\Psi(p,t)&=&|\tilde{\Psi}(p,t)|^2=\int^{\infty}_{-\infty} dx\; W(q,p;t).
\end{eqnarray}

\begin{figure}[ht!]
   \centering
   \subfloat[]{
        \label{fig1:museo:a}         
              \includegraphics[width=0.44\textwidth]{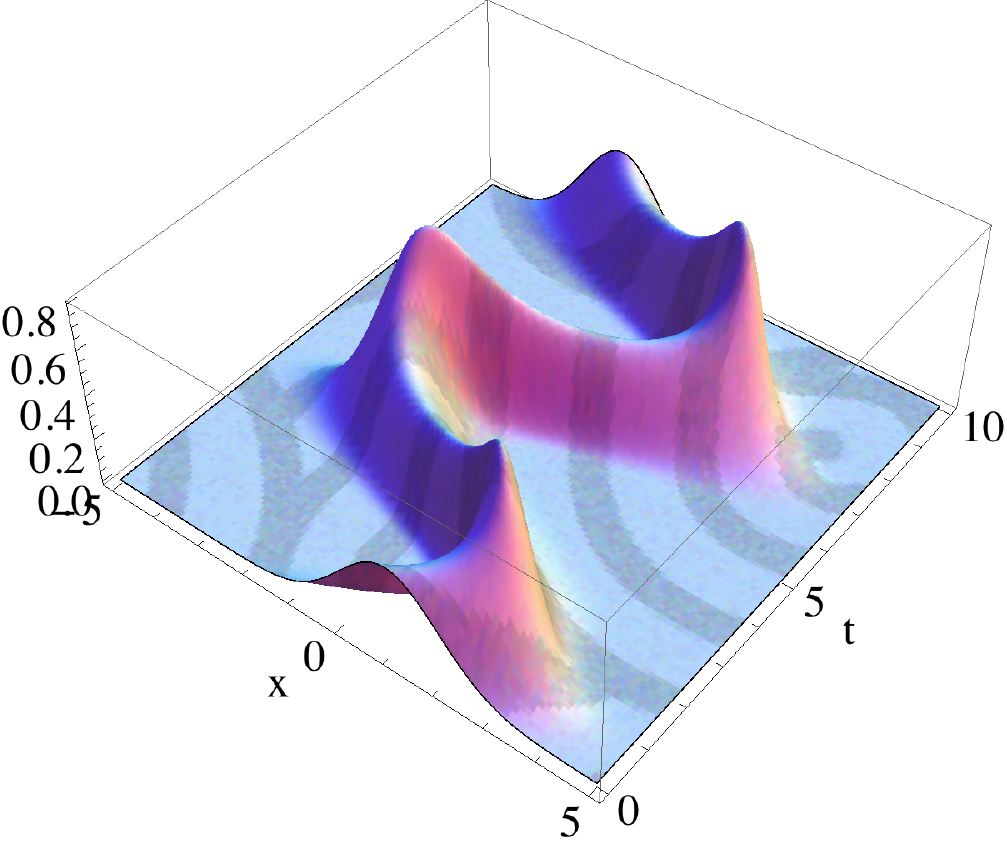}
              }
   \hspace{0.05\linewidth}
   \subfloat[]{
        \label{fig1:museo:b}         
                \includegraphics[width=0.44\textwidth]{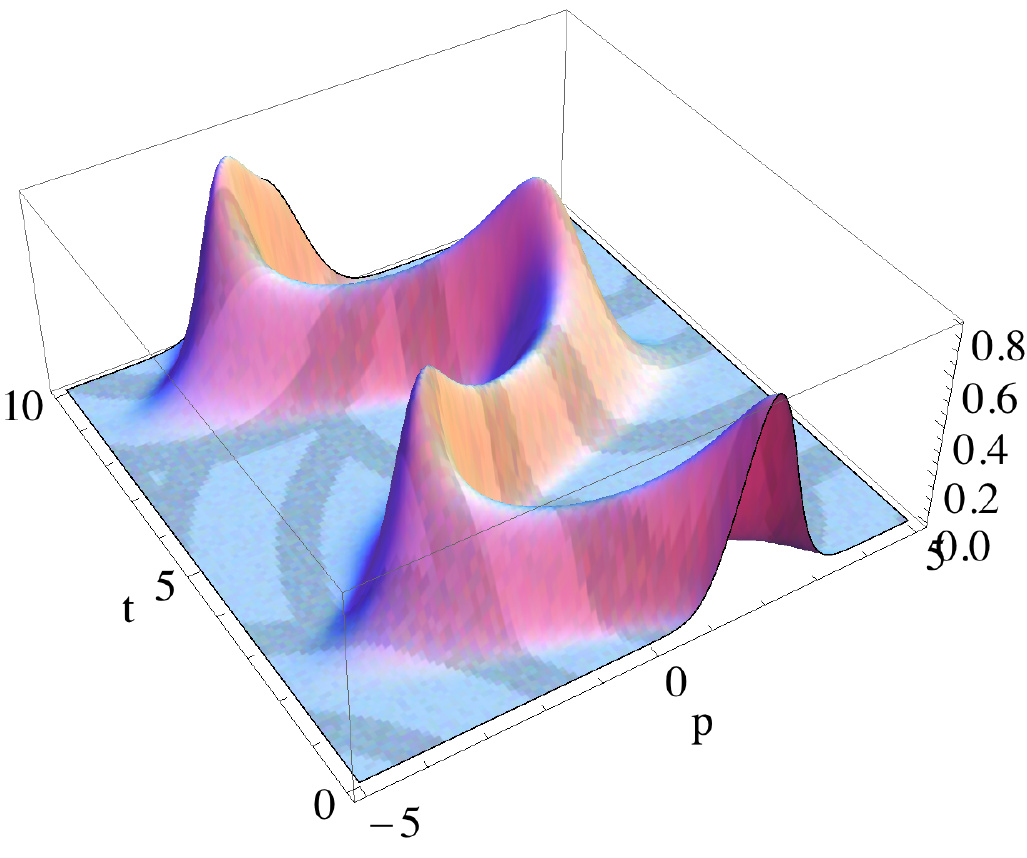}
                }
   \caption{\footnotesize (a) Probability distribution in position representation. (b) Probability distribution in momentum representation. }
   	\label{distributions}                
\end{figure}
\noindent
For the case under consideration, the probability distribution functions in position and momentum representations are plotted in Fig.~\ref{distributions}. Note that the maximum of the distribution follows the classical trajectory and its width changes periodically in time according to the values of $\sigma_x(t)$ and $\sigma_p(t)$.

Similar to the probability distributions $\Psi^\ast(x,t)\Psi(x,t)=|\Psi(x,t)|^2~=~\rho(x,t)$, the Wigner function also fulfils a continuity equation but now, as in the classical case, in phase space, i.e., 
\begin{equation}
\frac{\partial}{\partial t}W(x,p;t)+\frac{p}{m}\frac{\partial}{\partial x}W(x,p;t)-\frac{\partial V(x,t)}{\partial x}\frac{\partial}{\partial p} W(x,p;t)=0
\end{equation}
with $p=\tilde{p}+\langle p \rangle(t)$, $x=\tilde{x}+\langle x \rangle(t)$.

Inserting (\ref{Wigner-physic}) into this equation together with the TDHO potential one has at
most bilinear or quadratic expressions in terms of $\tilde{x}$ and $\tilde{p}$. The terms proportional to $\tilde{p}^2$ yield Eq. (\ref{sx}) in Appendix~\ref{Appendix-A}; those proportional to $\tilde{x}^2$ lead to Eq. (\ref{sp}) and those proportional to $\tilde{x}\tilde{p}$ to Eq. (\ref{sxp}). So, the set of three coupled first-order differential equations connecting the time-evolution of the uncertainties are regained in this way.

\section{ Conclusions and perspectives}

It has been shown that the information about the dynamics of the quantum systems under consideration is not only obtainable from the TDSE, but also from a complex Riccati equation. Solution of this equation directly provides the information about the time-dependence of the quantum mechanical uncertainties of position and momentum and their correlation. As the time-evolution of these quantities essentially determines observable quantum effects like tunnelling currents (see Appendix~\ref{Appendix-C}), the laws describing this evolution and the dependence on initial conditions are of major importance. The way the initial conditions enter these laws has been shown explicitly in Sections 2 and 3. Using the standard textbook example of the HO, it has been shown how a simple change of the initial conditions (WP with a width of the ground state Gaussian function vs. a width differing from this) can lead from a static situation (WP with constant width) to a dynamic one (WP with oscillating width). This is due to the fact that the static width refers only to a particular solution of the corresponding Riccati equation, whereas the oscillating width is related to the general solution. A possible second solution of the Riccati equation is discarded in the static case because it would lead to a diverging WP. In the case of the general solution, also this second particular solution can lead to physically-reasonable results (examples will be given in Part II which considers dissipative systems).

Furthermore, the connections between the classical degrees of freedom and the corresponding quantum uncertainties have been shown in detail. Particularly from the linear invariants (see Appendix~\ref{Appendix-B}), linearizing the Riccati equation (\ref{Riccati-equation}) to the complex Newtonian equation (\ref{Complex-Newton-equation}) together with the conservation law (\ref{conservation-law}) can show how quantum uncertainties can be expressed in terms of classical position and momentum and vice versa.

The real and imaginary parts of the complex Newtonian equation also allow for the definition of the propagator or Feynman kernel of the system as shown in (\ref{propagator}). In polar coordinates, the absolute value of $\lambda(t)$ is identical to the variable $\alpha(t)$ that transforms the complex Riccati equation into the real Ermakov equation and is directly proportional to the WP width. With the equation of motion for $\alpha(t)$ and the classical trajectory $\eta(t)$, it is possible to find a dynamical invariant, the Ermakov invariant (\ref{Ermakov-invariant}), that is still an invariant in cases where the Hamiltonian of the system is not (e.g., for an oscillator with TD frequency $\omega(t)$ or for certain dissipative systems as will be shown in Part III). Moreover, this invariant allows one to immediately write down the corresponding Wigner function of the system.

These results, obtained from standard quantum systems with exact analytical solutions, can however serve as the basis for generalizations. In Part II, it will be shown how the formalism can be extended to dissipative systems with linear velocity dependent friction forces. This will still be possible in the framework of analytical solutions. The next challenge will be systems (without and with dissipation) where no analytic solutions of the corresponding classical problem exist, i.e., Newtonian equations with potentials $V(x,t)$ that cannot be solved in a closed form. In these cases, it is reasonably assumed that Newtons equation still applies.

The same will be assumed for the Ermakov equation (corresponding to a complex Newtonian equation). As we know the connections between the classical variables and the corresponding quantum mechanical uncertainties (particularly the position uncertainty related to $\alpha (t)$), from the problems with analytical solutions discussed in this paper, it should be sufficient to calculate numerically the solutions for $\eta(t)$ and $\dot{\eta}(t)$ in these cases to derive the corresponding quantum uncertainties and properties like tunnelling currents and ground state energies. Besides, once $\eta(t)$, $\dot{\eta}(t)$, $\alpha(t)$ and $\dot{\alpha}(t)$ are available, also $\lambda(t)=\lambda_R(t)+i\lambda_I(t)$ can be ascertained and, thus, the parameters defining the propagator of the system and, via the Ermakov invariant (that might not be an invariant in these cases), the corresponding Wigner function determined.

That the values of quantum mechanical properties like tunnelling currents are sensitive to the choice of  initial conditions (and may change qualitatively according to this choice) can already be shown in analytic form considering dissipative systems with linear velocity dependent friction forces. This is dealt with in Part II.


\appendix

\section{Equations of motion of the quantum uncertainties}
\label{Appendix-A}

\numberwithin{equation}{section}
\setcounter{equation}{0}
\numberwithin{table}{section}
\setcounter{table}{0}

For the parametric oscillator (for $\frac{d}{dt} \omega (t) = 0$ this also applies for the HO), the uncertainties of position and momentum and their correlation satisfy the following closed set of first order differential equations
\begin{eqnarray}
\frac{d\sigma^2_x(t)}{dt}&=&\frac{2}{m}\sigma_{xp}(t),\label{sx}\\
\frac{d\sigma^2_p(t)}{dt}&=&-2m\omega^2(t)\sigma_{xp}(t),\label{sp}\\
\frac{d\sigma_{xp}(t)}{dt}&=&\frac{\sigma_p^2(t)}{m}-m\omega^2(t)\sigma^2_x(t).
\label{sxp}
\end{eqnarray}
This system has an invariant given by $I_{SR}=\sigma_x^2(t)\sigma_p^2(t)-\sigma_{xp}^2(t)$, corresponding to the Schr\"odinger--Robertson uncertainty product~\cite{26}.  

To solve the system defined by (\ref{sx}), (\ref{sp}) and (\ref{sxp}),  it can be written as a third-order ordinary differential equation for $\sigma_x^2(t)$, 
\begin{equation}
\frac{d^3\sigma^2_x(t)}{dt^3}+4\omega^2(t)\frac{d\sigma_x^2(t)}{dt}+4\omega(t)\frac{d\omega(t)}{dt}\sigma^2_x(t)=0.
\label{dif-third-ord}
\end{equation}
This equation is known as the normal form of maximal symmetry because all linear or linearizable third-order ordinary differential equation can be transformed into (\ref{dif-third-ord}),~\cite{27}. 

Inserting the expression 
\begin{equation}
\sigma^2_x(t)=\frac{\hbar}{2m} \alpha^2(t)
\label{sigmaq}
\end{equation}
 into (\ref{dif-third-ord}) shows that (\ref{sigmaq}) is only a solution of (\ref{dif-third-ord}) if $\alpha(t)$ is  a solution of the Ermakov equation (\ref{Ermakov-equation}). From (\ref{sigmaq}) and (\ref{sx}) the position-momentum uncertainty correlation $\sigma_{xp}(t)$ can be obtained immediately as
 \begin{equation}
\sigma_{xp}(t)=\frac{\hbar}{2}\dot{\alpha}(t)\alpha(t) 
\label{sigmaqp}
\end{equation}
and from $\sigma_{xp}(t)$ and (\ref{sxp}) it follows that 
\begin{equation}
\sigma^2_p(t)=\frac{m\hbar}{2}\left(\dot{\alpha}^2(t)+\frac{1}{\alpha^2(t)}\right) .
\label{sigmap}
\end{equation}


\section{Ermakov solutions and linear invariant operators}

\label{Appendix-B}

According to classical Hamiltonian mechanics, the integration of equations of motion becomes trivial if one can find a canonical transformation to variables that are constants. The same applies to the corresponding quantum mechanical operators. The linear TD operators $\hat{X}$ and $\hat{P}$ for the parametric oscillator are defined by the transformation
\begin{equation}
 \left(\begin{array}{c} 
      \hat{X} \\
      \hat{P} \\
   \end{array}\right)=
   \left(
      \begin{array}{cc} 
         g_1(t) & f_1(t) \\
         g_2(t) & f_2(t) \\
      \end{array}
   \right)
   \left(\begin{array}{c} 
      \hat{x} \\
      \hat{p} \\
   \end{array}\right)
   \label{linear-invariant}
\end{equation}
which satisfies the commutation relations of position and momentum operators, with the initial condition $\hat{X}(t_0)=\hat{x}$, $\hat{P}(t_0)=\hat{p}$. The time-invariance implies that the functions $g_i(t)$ and $f_i(t)$ satisfy the coupled system of differential equations
\begin{eqnarray}
\dot{g_i}(t)&=&m\omega^2(t) f_i(t),\\
\dot{f_i}(t)&=&-\frac{1}{m} g_i(t),
\end{eqnarray} 
$i=1,2$, which is equivalent to the second-order differential equation
\begin{equation}
\ddot{f}_i+\omega^2(t)f_i=0.
\end{equation}
With these invariant operators the most general quadratic invariant operator can be proposed as
\begin{equation}
\hat{I}=\frac{1}{2}\left[ A \hat{X}^2+B \hat{P}^2+C\left(\hat{X}\hat{P}+\hat{P}\hat{X}\right)\right] 
\label{quadratic-invariant}
\end{equation}
where $A$, $B$ and $C$ are constants still to be determined. As we already know a quadratic invariant for the system, the Ermakov invariant we assume that the invariant (\ref{quadratic-invariant}) is proportional to the operator corresponding to the Ermakov invariant,\footnote{The usual definition of the invariant is without the factor $\frac{m}{\hbar}$. The conventional expression is obtained for $m=\hbar=1$.},
\begin{equation}
\hat{I}=\frac{m}{2\hbar}\left[ \left( \alpha(t)\frac{\hat{p}}{m}-\dot{\alpha}(t)\hat{x}\right)^2 +\left(\frac{\hat{x}}{\alpha(t)}\right)^2\right].
\label{Erm-invariant}
\end{equation}
Taking Eq. (\ref{quadratic-invariant}) to be equal to (\ref{Erm-invariant}) by expressing $\hat{X}$ and $\hat{P}$ according to (\ref{linear-invariant}) in terms of $\hat{x}$, $\hat{p}$ and comparing the coefficients of $\hat{x}$, $\hat{p}$ and $\hat{x}\hat{p}+\hat{p}\hat{x}$, one obtains for $\alpha(t)$
\begin{equation}
\alpha(t)=\pm\sqrt{m\hbar}\left[A f_1^2(t)+B f_2^2(t) \mp2|C| f_1(t) f_2(t)\right]^{1/2},
\label{Solution-Ermakov-alpha0}
\end{equation}
with
\begin{equation}
|C|=\pm\sqrt{A B -\frac{1}{\hbar^2}}.
\end{equation}
Using the initial conditions for $f_1(t)$ and $f_2(t)$ and requiring that $\alpha_0\equiv\alpha(t_0)$ $\dot{\alpha}_0 \equiv \dot{\alpha}(t_0)$, one obtains
\begin{equation}
A =\frac{m}{\hbar} \Big(\dot{\alpha}_0^2 +\frac{1}{\alpha^2_0}\Big),\quad
B =\frac{\alpha_0^2}{m\hbar},\quad
|C| =\frac{\dot{\alpha}_0\alpha_0}{\hbar}.
\end{equation}
The coefficients $A$, $B$ and $|C|$ can be expressed in terms of the initial uncertainties and correlation function as
\begin{equation}
A =\frac{2}{\hbar^2}\sigma_{p_0}^2,\quad
B =\frac{2}{\hbar^2}\sigma_{x_0}^2,\quad
|C| =\frac{2}{\hbar^2}\sigma_{xp_0}.
\end{equation}

Obviously there is an ambiguity in sign in the expression for $\alpha (t)$. Considering the negative sign, the solution $\alpha(t)$ admits that $\alpha(t_0)$ and $\dot{\alpha}(t_0)$ may be different from zero. Considering the positive sign, $\alpha(t_0)$ is still different from zero, but $\dot{\alpha}(t_0)=0$. Therefore, in the more general first case, $\alpha(t)$ takes the form
\begin{equation}
\alpha(t)=\sqrt{\frac{2m}{\hbar}\left[\sigma_{p_0}^2 f_1^2(t)+\sigma_{x_0}^2f_2^2(t) \mp 2\sigma_{x_0p_0}f_1(t)f_2(t)\right]}.
\end{equation}


\section{Quantum uncertainties and tunnelling currents}

\label{Appendix-C}
 
The continuity equation describing the time-evolution of the probability distribution $\rho=\Psi^\ast(x,t)\Psi(x,t)=|\Psi(x,t)|^2$ in position space which corresponds to the TDSE (\ref{TDSE}) is given by 
\begin{equation}
\frac{\partial}{\partial t}\rho(x,t)+\frac{\partial}{\partial x}\left[ \rho(x,t)v_{-}(x,t) \right]=0,
\end{equation}
where the velocity field $v_{-}(x,t)$ is defined as
\begin{equation}
v_{-}(x,t)=\frac{\hbar}{2im}\left[ \frac{\frac{\partial}{\partial x}\Psi(x,t)}{\Psi(x,t)}-\frac{\frac{\partial}{\partial x}\Psi^\ast(x,t)}{\Psi^\ast(x,t)} \right]=\frac{\hbar}{2im}\frac{\partial}{\partial x}\ln\frac{\Psi(x,t)}{\Psi^\ast(x,t)}
\end{equation}
and the corresponding current is $\rho(x,t)v_{-}(x,t)$.

The velocity $v_{-}(x,t)$ obviously only depends on the phase of the wave function $\Psi(x,t)$ and, therefore in our case of Gaussian WPs with complex coefficient $C(t)$, also on $C_R(t)=\frac{\dot{\alpha}(t)}{\alpha(t)}$. For these WPs the velocity field can  be written explicitly as
\begin{equation}
v_{-}(x,t)=\dot{\eta}(t) + \frac{\dot{\alpha}(t)}{\alpha(t)}\tilde{x}
\end{equation}
which shows that, apart from the classical velocity $\dot{\eta}(t)$, also the tunnelling contribution depending on $\frac{\dot{\alpha}(t)}{\alpha(t)}$ is involved. Different initial conditions of the Riccati or Ermakov equation can lead to different time-evolution of $\alpha(t)$ and thus to different tunnelling currents.


\section{Uncertainties and correlations for the free motion}

\label{Appendix-D} 

In the case of the free motion ($V=0$) the results corresponding to Eqs. (\ref{alpha-HO}) and (\ref{sx-HO})-(\ref{sxp-HO}) can be obtained easily using 
\begin{equation}
\lim_{\omega_0 \to 0} \frac{\sin{\omega_0 t}}{\omega_0} = t, \quad \lim_{\omega_0 \to 0} \cos{\omega_0t} = 1.
\end{equation}

The solution of the Ermakov equation can be written as
\begin{eqnarray}
\alpha(t)&=&\sqrt{\frac{2m}{\hbar}\left[ \frac{\sigma^2_{p_0}}{m^2}t^2+\sigma^2_{x_0}\mp2\frac{\sigma_{xp_0}}{m}t \right]} \nonumber\\
&=&\alpha_0\sqrt{\left( 1 \pm \frac{\dot{\alpha}_0}{\alpha_0} t \right)^2+\frac{t^2}{\alpha^4_0}}
\end{eqnarray}
and the uncertainties and correlation functions take the form
\begin{eqnarray}
\sigma^2_x(t)&=&\frac{\sigma^2_{p_0}}{m^2} t^2 + \sigma^2_{x_0} \mp 2\frac{\sigma_{xp_0}}{m}t,\\
\sigma^2_p(t)&=&\sigma^2_{p_0},\\
\sigma_{xp}(t)&=&\frac{\sigma^2_{p_0}}{m} t + \sigma_{xp_0}.
\end{eqnarray}
For $\dot{\alpha}_0=0$, or $\sigma_{xp_0}=0$, the well-known textbook results are regained. If these initial values do not vanish, two additional solutions for $\sigma_x^2(t)$ are possible. However, one must carefully check if both mathematical solutions are also physically reasonable. For example, calculating $\sigma_{xp}(t)$ according to (\ref{sx}) by taking the derivative of $\sigma_x^2(t)$ would lead to $\sigma_{xp}(t)=\frac{\sigma^2_{p_0}}{m} t \mp \sigma_{xp_0}$. For $t=0$, this would also allow for $\sigma_{xp}(t_0)=-\sigma_{xp_0}$ which is obviously incorrect if $\sigma_{xp_0}\neq0$. So, only the plus-sign provides a physically reasonable solution corresponding to a physical system with non-vanishing initial correlations ($\dot{\alpha}_0\neq$ or $\sigma_{xp_0}\neq0$).


\subsection*{Acknowledgments}
This work was partially supported by CONACyT-M\'exico (under projects 238494 and 152574).


\end{document}